\documentclass[10pt,a4paper]{article}
\usepackage[latin1]{inputenc}
\usepackage{amsmath}
\usepackage{amsfonts}
\usepackage{amssymb}
\usepackage{graphicx}
\begin{document}
\title{Complex polynomials in engineering}
\maketitle
\tableofcontents
\section{Abstract}
Techniques for the evaluation of complex polynomials with one and two variables are introduced. Polynomials arise in may areas such as control systems, image and signal processing, coding theory, electrical networks, etc., and their evaluations are time consuming. This paper introduces new evaluation algorithms that are straightforward with fewer arithmetic operations. \\ \\
\textbf{Keywords:} Complex polynomial, one or two variables, control systems. \\
\section{Introduction}
Problems involving polynomial possess a long history in mathematics, but in the last few decades have attracted a lot of attention because of their importance in areas of contemporary applied mathematics, including control systems, electrical networks, image and signal processing, and coding theory. In the last decades, a tremendous improvements in microelectronics technology have led to an advancement in microprocessors which in their turn have increased the availability of low cost personal computers. The personal computers have had multiplicative effects on a number of areas such as control systems, signal processing, etc.. However, at the beginning, many of the personal computers had compilers without the capability of complex arithmetic. In this  paper, a mapping from the complex number domain to the skew symmetric domain is study and  used in combination with the Horner technique to develop  polynomial evaluation algorithms. These algorithms do not involve any complex arithmetic and require fewer floating point arithmetic operations than the conventional techniques. Furthermore, the algorithms can be used in the stability analysis of linear continuous time invariant systems. In the sequel, first we will study the mapping (isomorphism) from the complex number domain to the skew symmetric domain, and we will develop the algorithms and then use them in the arithmetic manipulations of complex polynomials and finally, we apply them to determine the stability of control systems. Following the same arguments as before, one can develop similar algorithms for the two variable polynomials.
\section{Notions of complex numbers}
Let $s=a+jb$ be a complex number belonging to the complex field $\mathbf{C}$, with $a$ as its real part and $b$ its imaginary part and $j=\sqrt{-1}.$ \\
Let  $R=\left[\begin{array}{cc}
a & b \\ 
-b & a
\end{array}  \right] $ be a rotation matrix belonging to the field of real skew symmetric matrices of rotation $\mathbf{R}$, with the properties:
\begin{itemize}
\item For $s=a+jb$, $R=\left[\begin{array}{cc}
Re\left\lbrace s \right\rbrace  & Im\left\lbrace s \right\rbrace \\ 
-Im \left\lbrace s \right\rbrace & Re\left\lbrace s \right\rbrace
\end{array}  \right]=\left[\begin{array}{cc}
a & b \\ 
-b & a
\end{array}  \right] $, where $Re\left\lbrace *\right\rbrace $ is the real part and $Im\left\lbrace *\right\rbrace $ is the imaginary part.
\item $R=aI+bJ$ where $I$ is the identity matrix and $J=\left[ \begin{array}{cc}
0 & 1 \\ 
-1 & 0
\end{array} \right]. $
\item $RR^{T}=R^{T}R=(a^{2}+b^{2})I$, $R$ is  orthogonal.
\item $R^{-1}=\frac{1}{a^{2}+b^{2}}R^{T}.$
\item Let $S=\frac{1}{\sqrt{a^{2}+b^{2}}}R,$ then $SJS^{T}=J$ and $S$ is a symplectic matrix.
\end{itemize}
The matrix $J$ has the following properties
\begin{itemize}
\item $J^{2}=-I$,
\item $J(-J)=I$,
\item $JJ^{T}=J(-J)=I$,
\item $JJ^{T}=J^{T}J=I.$
\item $J^{k}=\left\lbrace \begin{array}{ccc}
-I & \text{if} & k=4l+2=2(1+2l), \\ 
-J & \text{if} & k=4l+3=2(1+2l)+1, \\ 
I & \text{if} & k=4l+4=4(1+l), \\ 
J & \text{if} & k=4l+5=4(1+l)+1,
\end{array}\right. $ \\
for   $ l=0,\, 1,\, 2,\, 3, \ldots .$ 
\end{itemize}
Notice that $R$ can be represented \[ R=\lambda_{1}E_{1}+\lambda_{2}E_{2}, \] where $\lambda_{1}=a+jb$ and $\lambda_{2}=\lambda_{1}^{*}=a-jb$ are the eigenvalues of $R$, and \\
$E_{1}=\frac{1}{2}\left[\begin{array}{cc}
1 & -j \\ 
j & 1
\end{array}  \right]=\frac{1}{2}\left( I-jJ\right) $ and $E_{2}=E^{*}_{1}=\frac{1}{2}\left[\begin{array}{cc}
1 & j \\ 
-j & 1
\end{array}  \right]=\frac{1}{2}\left( I+jJ\right), $  are the spectral decomposition or skew Hermitian matrices. Therefore, for any known function $f(.)$, one has  \[ f(R)=f(\lambda_{1})E_{1}+f(\lambda_{2})E_{2}, \] and of  particular interest 
\begin{eqnarray*}
R^{k} & = & \lambda_{1}^{k}E_{1}+\lambda_{2}^{k}E_{2}, \\
& = & \sum_{m=0}^{k}\binom{k}{m} a^{m}b^{k-m}J^{k-m}.
\end{eqnarray*}
Assume that $b\neq 0$ and let $T$ be a similarity transformation matrix
\[ T=\left[ \begin{array}{cc}
-a & 1 \\
-b & 0 
\end{array} \right],  \]  such that  \[ J_{c}=T^{-1}JT=\frac{1}{b}\left[ \begin{array}{cc}
-a & 1 \\
-(a^{2}+b^{2}) & a 
\end{array} \right]  \] and \[ R_{c}=T^{-1}RT= \left[ \begin{array}{cc}
0 & 1 \\
-(a^{2}+b^{2}) & 2a
\end{array}\right].  \] Thus $J_{c}^{k}=T^{-1}J^{k}T$ and $R_{c}^{k}=T^{-1}R^{k}T$, for some integer $k$,
where
\[J_{c}^{k}=\left\lbrace \begin{array}{ccc}
-I & \text{if} & k=4l+2=2(1+2l), \\ 
-J_{c} & \text{if} & k=4l+3=2(1+2l)+1, \\ 
I & \text{if} & k=4l+4=4(1+l), \\ 
J_{c} & \text{if} & k=4l+5=4(1+l)+1.
\end{array}\right.\]

There is a mapping between the complex numbers $s$ and rotation matrices $R$, 
\[ s\longleftrightarrow R\] Such a map is one to one and onto, and  therefore an isomorphism. If $s_{1}=a_{1}+jb_{1}$ and $s_{2}=a_{2}+jb_{2}$ are two complex numbers in $\mathbf{C}$, then $R_{1}=a_{1}I+b_{1}J$ and $R_{2}=a_{2}I+b_{2}J$ are in $\mathbf{R}$, with the properties:
\begin{itemize}
\item $s_{1}+s_{2}=(a_{1}+a_{2})+j(b_{1}+b_{2})\in \mathbf{C}$, then \\ $R_{1}+R_{2}=(a_{1}+a_{2})I+(b_{1}+b_{2})J=Re\lbrace s_{1}+s_{2}\rbrace I+Im\lbrace s_{1}+s_{2}\rbrace J\in \mathbf{R}$,
\item $s_{1}-s_{2}=(a_{1}-a_{2})+j(b_{1}-b_{2})\in \mathbf{C}$, then \\ $R_{1}-R_{2}=(a_{1}-a_{2})I+(b_{1}-b_{2})J=Re\lbrace s_{1}-s_{2}\rbrace I+Im\lbrace s_{1}-s_{2}\rbrace J\in \mathbf{R}$,
\item $s_{1}*s_{2}=(a_{1}a_{2}-b_{1}b_{2})+j(a_{2}b_{1}+a_{1}b_{2})\in \mathbf{C}$, then \\ $R_{1}R_{2}=(a_{1}a_{2}-b_{1}b_{2})I+(a_{2}b_{1}+a_{1}b_{2})J=Re\lbrace s_{1}*s_{2}\rbrace I+Im\lbrace s_{1}*s_{2}\rbrace J\in \mathbf{R}$,
\item $\frac{s_{1}}{s_{2}}=\frac{a_{1}a_{2}+b_{1}b_{2}}{a_{2}^{2}+b_{2}^{2}}+j\frac{a_{2}b_{1}-a_{1}b_{2}}{a_{2}^{2}+b_{2}^{2}} \in \mathbf{C}$, then \\ $R_{1}R_{2}^{-1}=R_{2}^{-1}R_{1}= \frac{1}{a_{2}^{2}+b_{2}^{2}}\left[(a_{1}a_{2}+b_{1}b_{2})I+(a_{2}b_{1}-a_{1}b_{2})J \right]=Re\lbrace \frac{s_{1}}{s_{2}}\rbrace I+Im\lbrace \frac{s_{1}}{s_{2}}\rbrace J \in  \mathbf{R}$, for $a_{2}\neq 0$ and/or $b_{2}\neq 0.$

\end{itemize}

In the next section, an algorithm for evaluating a polynomial in a given point will be developed.
\section{Polynomials with real coefficients}
Suppose now that $p(s)$ is an $n^{th}$ order  polynomial with real coefficients, of the form \[ p(s)=\sum_{l=0}^{n}\alpha_{l}s^{n-l}=u_{p}+jv_{p}.\] Let $w_{0}=0$ such that $p(s)=w_{0}s^{n+1}+\sum_{l=0}^{n}\alpha_{l}s^{n-l}$ which can be written as
\[ p(s)=\underbrace{\alpha_{n}+s(\underbrace{\alpha_{n-1}+\cdots +s(\underbrace{\alpha_{2}+s(\underbrace{\alpha_{1}+s(\underbrace{\alpha_{0}+sw_{0}}_{w_{1}})}_{w_{2}})}_{w_{3}})\cdots)}_{w_{n}}}_{w_{n+1}}, \] using
Horner's algorithm, one gets
\[ \begin{array}{ccc}
w_{1} & = & sw_{0}+\alpha_{0}  \\
w_{2} & = & sw_{1}+\alpha_{1}  \\
\vdots & & \vdots \\
w_{n} & = & sw_{n-1}+\alpha_{n-1}  \\
w_{n+1} & = & sw_{n}+\alpha_{n}  = p(s).
\end{array} \]
In general, $w_{l+1}=sw_{l}+\alpha_{l}$ for $0\leq l\leq n$, with $w_{0}=0$ and $w_{n+1}=p(s).$

Since the mapping $s\longleftrightarrow R$ is an isomorphism, and 
\[ s^{l}\longleftrightarrow R^{l}=\left[ \begin{array}{cc}
Re\left\lbrace s^{l}\right\rbrace & Im\left\lbrace s^{l}\right\rbrace \\ 
-Im\left\lbrace s^{l}\right\rbrace & Re\left\lbrace s^{l}\right\rbrace
\end{array}\right]= Re\left\lbrace s^{l}\right\rbrace I+ Im\left\lbrace s^{l}\right\rbrace J\] then one has
\begin{eqnarray*}
p(s)\longleftrightarrow p(R) & = & \left[ \begin{array}{cc}
Re\left\lbrace p(s)\right\rbrace & Im\left\lbrace p(s)\right\rbrace \\ 
-Im\left\lbrace p(s)\right\rbrace & Re\left\lbrace p(s)\right\rbrace
\end{array}\right] \\
& = & Re\left\lbrace p(s)\right\rbrace I+ Im\left\lbrace p(s)\right\rbrace J \\ & = & \sum_{l=0}^{n}\alpha_{l}R^{n-l}. 
\end{eqnarray*}
The real and imaginary parts of the polynomial $p(s)$ can be determined by the following vector multiplication
\[ p(R)\left[ \begin{array}{c}
1 \\ 0
\end{array}\right]= \left[ \begin{array}{c}
Re\left\lbrace p(s)\right\rbrace \\ -Im\left\lbrace p(s)\right\rbrace \end{array} \right]. \]

Let the vector $w=\left[ \begin{array}{c}
0 \\ 1
\end{array}\right]$,  since $p(R_{c})=T^{-1}p(R)T$ or $Tp(R_{c})=p(R)T$ then the multiplication of both sides of the last expression by the vector $w$ yields $Tp(R_{c})w=p(R)Tw$, but $Tw=\left[ \begin{array}{cc}
-a & 1 \\
-b & 0 
\end{array} \right]\left[ \begin{array}{c}
0 \\ 1
\end{array}\right]=\left[ \begin{array}{c}
1 \\ 0
\end{array}\right]$. Therefore, 
\[ p(R)\left[ \begin{array}{c}
1 \\ 0
\end{array}\right]= \left[ \begin{array}{c}
Re\left\lbrace p(s)\right\rbrace \\ -Im\left\lbrace p(s)\right\rbrace \end{array} \right]=Tp(R_{c})w.  \] 
To determine the main equations of  a polynomial evaluation procedure at a particular point, let $z_{0}=\left[ \begin{array}{c}
0 \\ 0
\end{array}\right]$, then the polynomial matrix equation
\[ p(R_{c})=\sum_{l=0}^{n}\alpha_{l}R_{c}^{n-l}, \]can be rewritten as
\[ p(R_{c})w=R_{c}^{n+1}z_{0}+\sum_{l=0}^{n}\alpha_{l}R_{c}^{n-l}w. \] 
Using Horner's rules, but this time with a matrix as a variable, one gets
\[ z_{l+1}=R_{c}z_{l}+\alpha_{l}w,\; \;\; \text{for}\;\; 0\leq l\leq n. \]
Hence, $p(R_{c})w=z_{n+1}$ and 
\[ \left[ \begin{array}{c}
Re\left\lbrace p(s)\right\rbrace \\ -Im\left\lbrace p(s)\right\rbrace \end{array} \right]=Tp(R_{c})w =Tz_{n+1}.\]
\section{Polynomials with complex coefficients}
Suppose that the $n^{th}$ order polynomial $p(s)$ has now complex coefficients, it can then be written as follows
\begin{eqnarray*}
p(s) & = & \sum_{l=0}^{n}\gamma_{l}s^{n-l}=u_{p}+jv_{p} \\
& = & \sum_{l=0}^{n}(\alpha_{l}+j\beta_{l})s^{n-l}= \sum_{l=0}^{n}\alpha_{l}s^{n-l}+j\sum_{l=0}^{n}\beta_{l}s^{n-l} \\
& = & p_{\alpha}(s)+jp_{\beta}(s)=(u_{\alpha}+jv_{\alpha})+j(u_{\beta}+jv_{\beta}) \\
& = & (u_{\alpha}-v_{\beta})+j(v_{\alpha}+u_{\beta}).
\end{eqnarray*} 
From the mapping of $p(s)=p_{\alpha}(s)+jp_{\beta}(s)$ onto $p(R)=p_{\alpha}(R)+jp_{\beta}(R)$, one can get as it is done previously, the real and imaginary parts of $p_{\alpha}(s)$ and $p_{\beta}(s)$ as follows
\begin{eqnarray*}
p_{\alpha}(R)\left[ \begin{array}{c}
1 \\ 0
\end{array}\right] & = & \left[ \begin{array}{c}
Re\left\lbrace p_{\alpha}(s)\right\rbrace \\ -Im\left\lbrace p_{\alpha}(s)\right\rbrace \end{array} \right]=\left[ \begin{array}{c}
u_{\alpha}\\ -v_{\alpha}
\end{array}\right] \\ 
p_{\beta}(R)\left[ \begin{array}{c}
1 \\ 0
\end{array}\right] & = & \left[ \begin{array}{c}
Re\left\lbrace p_{\beta}(s)\right\rbrace \\ -Im\left\lbrace p_{\beta}(s)\right\rbrace \end{array} \right]=\left[ \begin{array}{c}
u_{\beta}\\ -v_{\beta}
\end{array}\right].  
\end{eqnarray*}
Since $R_{c}^{l}=T^{-1}R^{l}T$, then $p_{\alpha}(R_{c})T=Tp_{\alpha}(R)$, $p_{\beta}(R_{c})T=Tp_{\beta}(R)$ and $p(R_{c})T=Tp(R)$. 
We now define  $U=\left[ \begin{array}{cc}
1 & 0 \\ 0 & -1
\end{array}\right]$ and the exchange matrix $G= \left[ \begin{array}{cc}
0 & 1 \\ 1 & 0
\end{array}\right]$ such that $GUG=-U$, $GGU=U$, $ UGU=-G$, $UUG=G$, $UG=J$ and $GU=-J$. However,
\[ p(R)\left[ \begin{array}{c}
1 \\ 0
\end{array}\right]= \left[ \begin{array}{c}
u_{p} \\ -v_{p} \end{array} \right]=Tp(R_{c})w  \] and
\[ Tp(R_{c})w  = Tp_{\alpha}(R_{c})w+UGTp_{\beta}(R_{c})w=Tp_{\alpha}(R_{c})w+JTp_{\beta}(R_{c})w, \] where
$Tp_{\alpha}(R_{c})w=\left[ \begin{array}{c}
u_{\alpha} \\ -v_{\alpha} \end{array} \right]$ and $Tp_{\beta}(R_{c})w=\left[ \begin{array}{c}
u_{\beta} \\ -v_{\beta} \end{array} \right].$ 
To determine the main equations of the evaluation procedure of a polynomial with complex coefficients \cite{Khier}, let $z_{\alpha_{0}}=z_{\beta_{0}}=\left[ \begin{array}{c}
0 \\ 0
\end{array}\right]$, then the polynomial matrix equations
$p_{\alpha}(R_{c})=\sum_{l=0}^{n}\alpha_{l}R_{c}^{n-l}$ and $p_{\beta}(R_{c})=\sum_{l=0}^{n}\beta_{l}R_{c}^{n-l}$ can be rewritten as
\[ p_{\alpha}(R_{c})w=R_{c}^{n+1}z_{\alpha_{0}}+\sum_{l=0}^{n}\alpha_{l}R_{c}^{n-l}w \] and \[ p_{\beta}(R_{c})w=R_{c}^{n+1}z_{\beta_{0}}+\sum_{l=0}^{n}\beta_{l}R_{c}^{n-l}w .\]
Using again Horner's rules, one can have
\[ z_{\alpha_{l+1}}=R_{c}z_{\alpha_{l}}+\alpha_{l}w,\; \;\; \text{for}\;\; 0\leq l\leq n. \] Similarly, \[ z_{\beta_{l+1}}=R_{c}z_{\beta_{l}}+\beta_{l}w,\; \;\; \text{for}\;\; 0\leq l\leq n. \] 
Hence, $p_{\alpha}(R_{c})w=z_{\alpha_{n+1}}$, $p_{\beta}(R_{c})w=z_{\beta_{n+1}}$, 
\[ \left[ \begin{array}{c}
u_{\alpha} \\ -v_{\alpha} \end{array} \right]=Tp_{\alpha}(R_{c})w =Tz_{\alpha_{n+1}}\] and \[ \left[ \begin{array}{c}
u_{\beta} \\ -v_{\beta} \end{array} \right]=Tp_{\beta}(R_{c})w =Tz_{\beta_{n+1}}.\] Therefore the real and imaginary parts of $p(s)$ can now be found as 
\[ \left[ \begin{array}{c}
u_{p} \\ v_{p} \end{array}\right]=UTz_{\alpha_{n+1}}+GTz_{\beta_{n+1}}. \]

\section{Arithmetic of complex polynomials}
Let the matrix $V=\left[ \begin{array}{cc}
1 & -j \\ 1 & j
\end{array}\right]$ and the complex conjugate polynomial $p^{*}(s^{*})=u_{p}-jv_{p}$, then one can have the following polynomial vector
\[ \left[ \begin{array}{c}
p(s) \\ p^{*}(s^{*})
\end{array}\right] = V\left( Tz_{\alpha_{n+1}}+JTz_{\beta_{n+1}}\right). \]
Since the absolute of a complex polynomial is given as \[ \vert p(s)\vert =\sqrt{p(s) p^{*}(s^{*})} \] and \[ p(s) p^{*}(s^{*})=\vert p(s)\vert^{2} = \frac{1}{2}\left[ \begin{array}{cc}
p(s) & p^{*}(s^{*})
\end{array}\right]G\left[ \begin{array}{c}
p(s) \\ p^{*}(s^{*})
\end{array}\right], \] then 
\[ p(s) p^{*}(s^{*})= \frac{1}{2}\left( Tz_{\alpha_{n+1}}+JTz_{\beta_{n+1}}\right)^{T}V^{*}GV\left( Tz_{\alpha_{n+1}}+JTz_{\beta_{n+1}}\right).  \] We now define the following matrices needed in the algebraic manipulation of the computational algorithm, $A=\left[\begin{array}{cc}
a^{2}+b^{2} & -a \\ -a & 1
\end{array} \right]$  and $B=\left[\begin{array}{cc}
0 & b \\ -b & 0
\end{array} \right]$, thus the product of $p(s)$ by  its complex conjugate is given as 
\[ p(s) p^{*}(s^{*})= \vert p(s)\vert^{2} =\left[ \begin{array}{cc}
z_{\alpha_{n+1}}^{T} &  z_{\beta_{n+1}}^{T}
\end{array}\right]\left[ \begin{array}{cc}
A & B \\ -B & A
\end{array}\right] \left[ \begin{array}{c}
z_{\alpha_{n+1}}\\ z_{\beta_{n+1}}
\end{array}\right]. \] 
Using the previous results, one can easily find the reciprocal of $p(s)$ as 
\[\frac{1}{p(s)}=\frac{p^{*}(s^{*})}{\vert p(s)\vert^{2}}\]
Furthermore, the sum and the difference of $p(s)$ with  its complex conjugate are given as 
\[ p(s)+ p^{*}(s^{*})=2\left[\begin{array}{cc}
0 & 1
\end{array} \right]\left( Az_{\alpha_{n+1}}+Bz_{\beta_{n+1}}\right)  \] and \[ p(s)- p^{*}(s^{*})=j2\left[\begin{array}{cc}
0 & 1
\end{array} \right]\left( -Bz_{\alpha_{n+1}}+Az_{\beta_{n+1}}\right)  ,\] respectively.

\section{Derivative of a polynomial}
Since the $n^{th}$ order polynomial $p(s)$ is an entire function, it is analytic everywhere, then its derivative exists and can be computed using similar algorithms as the previous ones: \\
Note that $p(s)=\sum_{l=0}^{n}\alpha_{l}s^{n-l}=u_{p}+jv_{p},$ then 

\begin{eqnarray*}
\frac{dp(s)}{ds}) & = & \sum_{l=0}^{n-1}\alpha_{l}(n-l)s^{n-l-1} \\
& = & \sum_{l=0}^{n-1}\nu_{l}s^{n-l-1} \\
 & = & \frac{du_{p}(s)}{ds}+j\frac{dv_{p}(s)}{ds}.
\end{eqnarray*}
Where $\nu_{l}=(n-l)\alpha_{l}.$ \\
Let $\dot{p}_{\nu}(s)=\frac{dp(s)}{ds})$, $u_{\nu}(s)=\frac{du_{p}(s)}{ds}$ and $v_{\nu}(s)=\frac{dv_{p}(s)}{ds},$
thus, if $\dot{p}_{\nu}(s)$ has
\begin{itemize}
\item real coefficients, then
\[ \left[ \begin{array}{c}
u_{\nu} \\ v_{\nu} \end{array}\right]=UTz_{\alpha_{n}}, \] where in this case 
\[ z_{\alpha_{l+1}}=R_{c}z_{\alpha_{l}}+(n-l)\alpha_{l}w,\; \;\; \text{for}\;\; 0\leq l\leq n-1. \]
\item complex coefficients
\[ \left[ \begin{array}{c}
u_{\nu} \\ v_{\nu} \end{array}\right]=UTz_{\alpha_{n+1}}+GTz_{\beta_{n+1}}, \] where in addition to $z_{\alpha_{l}}$ computed above, we also have
\[ z_{\beta_{l+1}}=R_{c}z_{\beta_{l}}+(n-l)\beta_{l}w,\; \;\; \text{for}\;\; 0\leq l\leq n. \] 
\end{itemize}
In the next section, the developed algorithms will be used in the computation of the control system frequency response needed the stability tests.

\section{Applications}
Assume that $s=j\omega$ or $a=0$ and $b=\omega$, and consider the following two cases:
\begin{enumerate}
\item the polynomial $p(s)$ has complex coefficients, then its real and imaginary parts are obtained as
\begin{equation}
\left[\begin{array}{c}
u_{p} \\v_{p}
\end{array} \right]=\left[\begin{array}{cc}
0 & 1 \\ \omega & 0
\end{array} \right]z_{\alpha_{n+1}} + \left[\begin{array}{cc}
-\omega & 0 \\ 0 & 1
\end{array} \right]z_{\beta_{n+1}}, 
\end{equation}
where 
\begin{equation}
z_{\alpha_{l+1}}=\left[\begin{array}{cc}
0 & 1 \\ -\omega^{2} & 0
\end{array} \right]z_{\alpha_{l}}+\alpha_{l}w 
\end{equation} 
and 
\begin{equation}
z_{\beta_{l+1}}=\left[\begin{array}{cc}
0 & 1 \\ -\omega^{2} & 0
\end{array} \right]z_{\beta_{l}}+\beta_{l}w 
\end{equation}
for  $0\leq l \leq n.$
Therefore the product and sum of $p(j\omega)$ with  its complex conjugate are determined as follows
\[ p(j\omega) p^{*}(-j\omega)=\vert p(j\omega)\vert^{2} = \left[ \begin{array}{cc}
z_{\alpha_{n+1}}^{T} &  z_{\beta_{n+1}}^{T}
\end{array}\right]\left[ \begin{array}{cccc}
\omega ^{2}& 0 & 0 & \omega \\ 0 & 1 & -\omega & 0 \\
0 & -\omega & \omega^{2} & 0 \\ \omega & 0 & 0 & 1
\end{array}\right] \left[ \begin{array}{c}
z_{\alpha_{n+1}}\\ z_{\beta_{n+1}}
\end{array}\right], \]  and 
\[ p(j\omega)+ p^{*}(-j\omega)=2\left(\left[\begin{array}{cc}
0 & 1
\end{array} \right]z_{\alpha_{n+1}}+\left[\begin{array}{cc}
-\omega & 1
\end{array} \right]z_{\beta_{n+1}}\right).  \] 
\item the polynomial $p(s)$ has real coefficients, then $z_{\beta_{l}}=0$ for  $0\leq l\leq n$, and its real and imaginary parts are determined as
\begin{equation}
\left[\begin{array}{c}
u_{p} \\v_{p}
\end{array} \right]=\left[\begin{array}{cc}
0 & 1 \\ \omega & 0
\end{array} \right]z_{\alpha_{n+1}},
\end{equation}
where 
\begin{equation}
z_{\alpha_{l+1}}=\left[\begin{array}{cc}
0 & 1 \\ -\omega^{2} & 0
\end{array} \right]z_{\alpha_{l}}+\alpha_{l}w  
\end{equation}
and
\[ p(j\omega) p^{*}(-j\omega)= 
z_{\alpha_{n+1}}^{T} \left[ \begin{array}{cc}
\omega ^{2}& 0   \\ 0 & 1 
\end{array}\right] 
z_{\alpha_{n+1}}\]  and 
\[ p(j\omega)+ p^{*}(-j\omega)=\left[\begin{array}{cc}
0 & 2
\end{array} \right]z_{\alpha_{n+1}}.  \] 

\end{enumerate}
In filter design and control systems theory  \cite{Chen}, the frequency response of a linear continuous time invariant system is often obtained by substituting $s=j\omega$ in the system transfer function $H(s)=\frac{q(s)}{p(s)},$ as the frequency is varied between two fixed values, then $q(j\omega)=Re\left\lbrace q(j\omega)\right\rbrace +jIm\left\lbrace q(j\omega)\right\rbrace$, $p(j\omega)=Re\left\lbrace p(j\omega)\right\rbrace +jIm\left\lbrace p(j\omega)\right\rbrace $ and $H(j\omega)=\vert H(j\omega)\vert e^{j\phi(\omega)}$ where the magnitude and phase of the system are given as
\[ \vert H(j\omega)\vert^{2} = H(j\omega) H^{*}(-j\omega)= \frac{(Re\left\lbrace q(j\omega)\right\rbrace)^{2} +(Im\left\lbrace q(j\omega)\right\rbrace)^{2}}{(Re\left\lbrace p(j\omega)\right\rbrace)^{2} +(Im\left\lbrace p(j\omega)\right\rbrace)^{2}} \] and 
\[ \phi(\omega)=\tan^{-1} \left( \frac{Re\left\lbrace p(j\omega)\right\rbrace Im\left\lbrace q(j\omega)\right\rbrace-Re\left\lbrace q(j\omega)\right\rbrace Im\left\lbrace p(j\omega)\right\rbrace}{Re\left\lbrace p(j\omega)\right\rbrace Re\left\lbrace q(j\omega)\right\rbrace+Im\left\lbrace q(j\omega)\right\rbrace Im\left\lbrace p(j\omega)\right\rbrace}\right) . \]
Therefore, the real and the imaginary parts of the system frequency response are
\[ Re\left\lbrace H(j\omega)\right\rbrace =\vert H(j\omega)\vert cos(\phi(\omega))\]  and 
\[ Im\left\lbrace H(j\omega)\right\rbrace =\vert H(j\omega)\vert sin(\phi(\omega)).\]
The calculation of the real and imaginary parts of the system frequency response using the expressions () and () takes $2(n+4)$ real multiplications and $2(n+1)$ real additions/subtractions for each value of the frequency. However, the conventional method requires about $6(n+1)+2$ real multiplications and $2(n+1)$ real additions/subtractions. For stability analysis, it is usually enough to have a rough sketch of the frequency response, since only the intersects on the real axis of the complex plane are to be calculated. However, for complex systems and system designs, the computation may be too tedious to handle and time consuming and sometimes, it may be necessary to know more about the plot itself before the final sketch is made correctly. The equations () and () provide the frequency response magnitude and phase to be used for the Bode plots drawing. Similarly, the equations () and () furnish the frequency response real and imaginary parts to be use in Nyquist plot drawing. Furthermore, the frequency response real and imaginary parts can also be used in the product computation of a rational function with its complex conjugate which is often used in optimal control to determine the symmetric or square root locus and to evaluate the cost function of a time invariant continuous system. \\ Note that if a numerator and/or denominator of a transfer function are given as products of first order factors, as
\[G(s)=\frac{\prod_{l=1}^{m}(1+T_{nl}s)}{\prod_{k=1}^{n}(1+T_{nk}s)},\;\;\; \text{with}\;\;\; n\geq m.\]
Where $T_{n1},\; \ldots ,\; T_{nm}$ and $T_{d1},\; \ldots ,\; T_{dn}$ are the system numerator and denominator positive constants respectively. The previous expression of the system transfer function can also be rewritten as
\[G(s)=K\frac{\sum_{l=0}^{m}q_{l}s^{m-l}}{\sum_{k=0}^{n}p_{k}s^{n-k}},\] where the gain $K$ is found to be given as \[K=\frac{\prod_{l=1}^{m}T_{nl}}{\prod_{k=1}^{n}T_{nk}},\] and the numerator and denominator coefficients $q_{01},\; \ldots ,\; q_{m}$ and $p_{0},\; \ldots ,\; p_{n}$ respectively, are computed as follows
\[c_{f}=(-1)^{f}\sum_{k=1}^{g}\prod_{l=1}^{f}r_{t}.\] Where \[g=\frac{n!}{f!(n-f)!}\]  and
\[t=\left\lbrace \begin{array}{cc}
(k+l-1)\, mod\, n & \text{}\;\; k+l\neq en+1, \\ 
n & \text{}\;\; k+l= en+1,
\end{array}\right.  \] for some integer $e.$ The roots and the coefficients of the numerator and denominator of $G(s)$ are then given respectively as
\begin{itemize}
\item Numerator   \[r_{f}=-\frac{1}{T_{nf}},\;\;\;\; q_{f}=c_{f},\;\;\;\; \text{for}\;\; f=1,\, \ldots , \, m.\]
\item Denominator   \[r_{f}=-\frac{1}{T_{df}},\;\;\;\; p_{f}=c_{f},\;\;\;\; \text{for}\;\; f=1,\, \ldots , \, n.\]
\end{itemize}
The system dynamic is usually described by its state space representation
\begin{eqnarray}
\dot{x}(t) & = & Ax(t)+Bu(t) \\
y(t) & = & Cx(t)
\end{eqnarray}
where $A$ is the $n\times n$ state matrix, $B$ is the $n\times 1$ input vector and $C$ is the $1\times n$ output vector. The system transfer function is 
\begin{eqnarray}
G(s) & = & C(sI-A)^{-1}B, \\ 
& = &\frac{1}{p(s)}Cadj(sI-A)B, \\
& = & \frac{q(s)}{p(s)}, 
\end{eqnarray}
where $p(s)=\sum_{l=0}^{n}p_{l}s^{n-l}$ is the system characteristic equation and $q(s)=Cadj(sI-A)B=\sum_{l=1}^{n}q_{l}s^{n-l}.$ The matrix adjoint is given as 
\begin{equation}
adj(sI-A)=\sum_{l=1}^{n}F_{l}s^{n-l} 
\end{equation}
where the matrices $F_{l}'s$ and the coefficients $p_{l}$ are computed using the Leverrier - Faddeva algorithm
\begin{eqnarray}
F_{l+1} & = & AF_{l}+p_{l}I, \\
p_{l} & = & -\frac{1}{l}tr(AF_{l}),
\end{eqnarray}
with $F_{1}=I_{n}$, $F_{n+1}=0$, $p_{0}=1$ and $tr(.)$ is the trace of a matrix. The matrices $F_{l}'s$ and the coefficients $p_{l}$ can also be given by
\begin{eqnarray}
F_{k} & = &\sum_{l=0}^{k-1}p_{l}A^{k-l-1}, \\
p_{k} & = &-\frac{1}{k}\sum_{l=0}^{k-1}p_{l}tr(A^{k-l}),
\end{eqnarray}
for $0\leq k\leq n.$ Thus, the transfer function numerator coefficients are found to be 
\begin{equation}
q_{k} =CF_{k}B=\sum_{l=0}^{k-1}p_{l}CA^{k-l-1}B.
\end{equation}

The matrix exponentiation is very important subject in control and signal processing.
A technique of conversion from a binary number to an integer is used to develop a fast algorithm for the computation of a matrix exponentiation. The method of conversion is given as follows: \\
Let $\rho_{0},\rho_{1},\ldots , \rho_{k-1}$ be the binary digits of the integer $\rho$, such  that
\begin{equation}
\left(\rho_{k-1}\rho_{k-2}\cdots \rho_{1}\rho_{0} \right)_{2}= \left( \rho\right) _{10} 
\end{equation}
To begin with, one must discard the  most significant digit $\rho_{k-1}$, and replace every other digit remaining that follows starting from the left, by a multiplication of 2, if $\rho_{i}=0$ and by a multiplication of 2 plus 1, if $\rho_{i}=1$. The most significant digit remaining is associated with the most inner parenthesis of the expansion. For example $117_{10}=1110101_{2}$ gives
$$117=\underbrace{1+2(\underbrace{2(\underbrace{1+2(\underbrace{2(\underbrace{1+2
(\underbrace{1+2}_{1})}_{1})}_{0})}_{1})}_{0})}_{1}$$
Based on the binary conversion given above, a fast algorithm for the computation of a matrix to a certain power is presented and it is given as follows: \\
To compute $A^{\rho}$ where $\left( \rho\right) _{10}=\left(\rho_{k-1}\rho_{k-2}\cdots \rho_{1}\rho_{0} \right)_{2}.$
\\
Let 
\begin{equation}
m_{k-i-1}\equiv \left\lbrace \begin{array}{cc}
m_{k-i}^{2} & \text{if}\;\; \rho_{k-i-1}=0 \\
m_{k-i}^{2}\; A& \text{if}\;\; \rho_{k-i-1}=1
\end{array}\right. 
\end{equation}
for $1\leq i\leq k-1$, where $m_{k-1}=A$ and $m_{0}=A^{\rho}.$

\section{Two dimensional polynomials}
The algorithms given in \cite{Khier} and repeated in the previous sections will be extended to the  two dimensional case. 
Although, a vast quantity of results has been found for the one variable polynomial, relatively fewer of these results can be extended to the two variables case. The most serious problem in the generalization of the one variable techniques to the two variable counterpart, is the fact that there is no fundamental theorem of algebra for polynomials in two independent variables. \\
Consider the linear continuous time invariant two dimensional (2D) system  described by the transfer function
\[G(s_{1}, s_{2})=\frac{q(s_{1}, s_{2})}{p(s_{1}, s_{2})},\]  where 
\begin{equation}
q(s_{1},s_{2})=\sum_{l=0}^{n_{1}}\sum_{k=0}^{m_{1}}q_{lk}s_{1}^{n_{1}-l}s_{2}^{m_{1}-k}=S_{1}QS_{2}^{T},
\end{equation}
the system characteristic polynomial is given as
\begin{equation}
p(s_{1},s_{2})=\sum_{l=0}^{n_{2}}\sum_{k=0}^{m_{2}}p_{lk}s_{1}^{n_{2}-l}s_{2}^{m_{2}-k}=S_{1}PS_{2}^{T},
\end{equation}
with
\[S_{1}=\left[\begin{array}{cccc}
s_{1}^{n_{2}} & s_{1}^{n_{2}-1} & \ldots & 1
\end{array} \right],\quad  S_{2}=\left[\begin{array}{cccc}
s_{2}^{m_{2}} & s_{2}^{m_{2}-1} & \ldots & 1
\end{array} \right],\]
\[Q =\left[\begin{array}{cccc}
q_{00} & q_{01} & \cdots & q_{0,m_{1}} \\ 
q_{10} & q_{11} & \cdots & q_{1,m_{1}} \\ 
\vdots & \vdots &  & \vdots \\ 
q_{n_{1},0} & q_{n_{1},1} & \cdots & q_{n_{1},m_{1}}
\end{array}  \right] \] and
\[P =\left[\begin{array}{cccc}
p_{00} & p_{01} & \cdots & p_{0,m_{2}} \\ 
p_{10} & p_{11} & \cdots & p_{1,m_{2}} \\ 
\vdots & \vdots &  & \vdots \\ 
p_{n_{2},0} & p_{n_{2},1} & \cdots & p_{n_{2},m_{2}}
\end{array}  \right]. \] The variables $s_{1}$ and $s_{2}$ are the 2D polynomial variables along the horizontal and vertical directions, respectively. \\
If $p(s_{1},s_{2})$ is separable (factorable), that is $p(s_{1},s_{2})=p_{1}(s_{1})p_{2}(s_{2})$ and the matrix $P$ has unit rank, then
\[P=\left[\begin{array}{c}
\eta_{0} \\ 
\eta_{1} \\ 
\vdots \\ 
\eta_{n_{2}}
\end{array}  \right] \left[\begin{array}{cccc}
\varrho_{0} & \varrho_{1} & \ldots &\varrho_{m_{2}}
\end{array}  \right] ,\] where $\eta_{l}=gcd\lbrace p_{lk}\rbrace ,$  $\varrho_{k}=gcd\lbrace p_{lk}\rbrace ,$  for $(0,0)\leq (l, k) \leq (n_{2}, m_{2}),$ then $p_{1}(s_{1})=\sum_{l=0}^{n_{2}}\eta_{l}s_{1}^{n_{2}-l}$ and $p_{2}(s_{2})=\sum_{k=0}^{m_{2}}\varrho_{k}s_{2}^{m_{2}-k}.$  \\ The algorithms developed in [] and repeated in section 4, can be used twice in two different directions horizontally and vertically to evaluate the 2D polynomial $p(s_{1},s_{2})$, whose magnitude would be the product of the individual magnitudes (horizontal and vertical), and its phase would be the sum of the individual phases. \\ If $p(s_{1},s_{2})$ cannot be factored, then let \begin{equation}
p(s_{1},s_{2})=\sum_{k=0}^{m_{2}}f_{k}(s_{1})s_{2}^{m_{2}-k},
\end{equation}
where
\begin{equation}
f_{k}(s_{1})=\sum_{l=0}^{n_{2}}p_{lk}s_{1}^{n_{2}-1}.
\end{equation}

Now, let $s_{1}=a_{1}+jb_{1}$ and $s_{2}=a_{2}+jb_{2}$ such that 
\begin{equation}
f_{k}(a_{1}+jb_{1})=\alpha_{k}+j\beta_{k}
\end{equation}
for $0\leq k\leq m_{2},$ and  $e_{0}=\left[ \begin{array}{c}
0 \\ 1
\end{array}\right]$, \[ R_{c1}=\left[ \begin{array}{cc}
0 & 1 \\
-(a_{1}^{2}+b_{1}^{2}) & 2a_{1}
\end{array}\right],\quad  T_{1}=\left[ \begin{array}{cc}
-a_{1} & 1 \\
b_{1} & 0 
\end{array} \right], \]
\[ R_{c2}=\left[ \begin{array}{cc}
0 & 1 \\
-(a_{2}^{2}+b_{2}^{2}) & 2a_{2}
\end{array}\right],\quad  T_{2}=\left[ \begin{array}{cc}
-a_{2} & 1 \\
b_{2} & 0 
\end{array} \right], \] and let $w_{0k}=\left[ \begin{array}{c}
0 \\ 0
\end{array}\right],$ then \begin{equation}
w_{l+1,k}=R_{c1}w_{lk}+p_{lk}e,
\end{equation}
for $0\leq l\leq n_{2},$ and 
\begin{equation}
\left[ \begin{array}{c}
\alpha_{k} \\ \beta_{k}
\end{array}\right]=T_{1}w_{n_{2}+1,k}.
\end{equation}
Hence, \[
p(a_{1}+jb_{1},s_{2})=\sum_{k=0}^{m_{2}}\alpha_{k}s_{2}^{m_{2}-k}+j\sum_{k=0}^{m_{2}}\beta_{k}s_{2}^{m_{2}-k}
\]
and \[ p(a_{1}+jb_{1},a_{2}+jb_{2})= \eta_{p}+\vartheta_{p}. \]
To compute $\eta_{p}$ and $\vartheta_{p}$, let $u_{0}=\left[ \begin{array}{c}
0 \\ 0
\end{array}\right]$, and $v_{0}=\left[ \begin{array}{c}
0 \\ 0
\end{array}\right]$,
\[u_{k+1}=R_{c2}u_{k}+\alpha_{k}e,\]  and \[v_{k+1}=R_{c2}v_{k}+\beta_{k}e.\]  The substitution of the expressions for $\alpha_{k}$ and  $\beta_{k}$, obtained from equation (10), into the previous two equations yields 
\begin{equation}
u_{k+1}=R_{c2}u_{k}+\left[ \begin{array}{cc}
-a_{1} & 1
\end{array}\right]w_{n_{2}+1,k}e,
\end{equation}
and 
\begin{equation}
v_{k+1}=R_{c2}v_{k}+\left[ \begin{array}{cc}
b_{1} & 0
\end{array}\right]w_{n_{2}+1,k}e.
\end{equation}
Thus,
\begin{equation}
\left[ \begin{array}{c}
\eta_{p} \\ \vartheta_{p}
\end{array}\right]=T_{2}u_{m_{2}+1}+J^{T}T_{2}v_{m_{2}+1}
\end{equation}
The same procedure can  be applied to the numerator of $G(s_{1}, s_{2})$  to obtain
\[q(a_{1}+jb_{1}, a_{2}+jb_{2})=\eta_{q}+j\vartheta_{q}.\]
The algorithm will established using equations (6) - (13). \\
Since \[G(s_{1}, s_{2})=\frac{q(s_{1}, s_{2})}{p(s_{1}, s_{2})},\] and \[G(a_{1}+jb_{1}, a_{2}+jb_{2})=\vert G(a_{1}+jb_{1}, a_{2}+jb_{2})\vert e^{j\phi_{h}(a_{1}+jb_{1}, a_{2}+jb_{2})}.\]
Let  $\phi_{p}(a_{1}+jb_{1}, a_{2}+jb_{2})=tan^{-1}\left(\frac{\vartheta_{p}}{\eta_{p}} \right),\;\;\; and\;\;\; \phi_{q}(a_{1}+jb_{1}, a_{2}+jb_{2})=tan^{-1}\left(\frac{\vartheta_{q}}{\eta_{q}} \right),$ then 
\begin{eqnarray}
\phi_{h}(a_{1}+jb_{1}, a_{2}+jb_{2}) & = & \phi_{q}(a_{1}+jb_{1}, a_{2}+jb_{2})-\phi_{p}(a_{1}+jb_{1}, a_{2}+jb_{2}) \nonumber \\
& = & tan^{-1}\left(\frac{\eta_{p}\vartheta_{q}-\eta_{q}\vartheta_{p}}{\vartheta_{p}\vartheta_{q}+\eta_{p}\eta_{q}} \right) 
\end{eqnarray}
and \[\vert G(a_{1}+jb_{1}, a_{2}+jb_{2})\vert =\sqrt{\frac{\eta_{q}^{2}+\vartheta_{q}^{2}}{\eta_{p}^{2}+\vartheta_{p}^{2}}}.\]
Therefore,
\[Re\lbrace G(a_{1}+jb_{1}, a_{2}+jb_{2})\rbrace =\vert G(a_{1}+jb_{1}, a_{2}+jb_{2})\vert cos\left( \phi_{h}(a_{1}+jb_{1}, a_{2}+jb_{2})\right),  \]
and
\[Im\lbrace G(a_{1}+jb_{1}, a_{2}+jb_{2})\rbrace =\vert G(a_{1}+jb_{1}, a_{2}+jb_{2})\vert sin\left( \phi_{h}(a_{1}+jb_{1}, a_{2}+jb_{2})\right).  \]

\section{Conclusion}
In this paper, we have introduced algorithms for the evaluation of complex polynomials with one and two variables and shown their applications to the stability analysis of linear continuous time invariant systems. These algorithms require fewer arithmetic operations than the existing techniques.

\end{document}